# Intersubband transitions in nonpolar GaN/Al(Ga)N heterostructures in the short- and mid-wavelength infrared regions

C. B. Lim[1,2], M. Beeler[1,2], A. Ajay[1,2], J. Lähnemann[1,2], E. Bellet-Amalric[1,2], C. Bougerol[1,3], and E. Monroy[1,2]

[1] *University Grenoble-Alpes, 38000 Grenoble, France*
[2] *CEA, INAC-SP2M, 17 av. des Martyrs, 38000 Grenoble, France*
[3] *CNRS, Institut Néel, 25 av. des Martyrs, 38000 Grenoble, France*

**ABSTRACT**

This paper assesses nonpolar *m*- and *a*-plane GaN/Al(Ga)N multi-quantum-wells grown on bulk GaN for intersubband optoelectronics in the short- and mid-wavelength infrared ranges. The characterization results are compared to those for reference samples grown on the polar *c*-plane, and are verified by self-consistent Schrödinger-Poisson calculations. The best results in terms of mosaicity, surface roughness, photoluminescence linewidth and intensity, as well as intersubband absorption are obtained from *m*-plane structures, which display room-temperature intersubband absorption in the range from 1.5 to 2.9 μm. Based on these results, a series of *m*-plane GaN/AlGaN multi-quantum-wells were designed to determine the accessible spectral range in the mid-infrared. These samples exhibit tunable room-temperature intersubband absorption from 4.0 to 5.8 μm, the long-wavelength limit being set by the absorption associated with the second order of the Reststrahlen band in the GaN substrates.

**I. INTRODUCTION**

GaN/AlGaN nanostructures have recently emerged as promising materials for new intersubband (ISB) devices covering a large portion of the infrared spectrum.[1–3] Their large conduction band offsets and sub-picosecond ISB relaxation times make them appealing for ultrafast photonics devices operating at telecommunication wavelengths.[4,5] Additionally, the large energy of the longitudinal-optical phonon in GaN (92 meV, 13 µm) opens prospects for room temperature THz lasers.[6,7]

So far, studies on ISB transitions in group-III-nitride multi-quantum-wells (MQWs) have mostly focused on polar *c*-plane structures. However, this crystallographic orientation comes with the complicating factor of a polarization-induced internal electric field, resulting in an asymmetric triangular potential in the quantum wells (QWs). The electric field renders ISB transition energies more sensitive to the strain state of the QWs,[8] and hampers the extension of ISB transitions towards far-infrared wavelengths. This quantum-confined Stark effect is a major hurdle for device design, although it has been partially compensated by the implementation of more complex step-QW designs.[9–12] The use of nonpolar *a* or *m* crystallographic orientations allows for GaN/Al(Ga)N systems to operate without the influence of this electric field[13] and facilitates the device design while still maintaining the benefits of GaN.

Regarding nonpolar materials, ISB optical absorption at λ ~ 2.1 µm has been reported in 1.75-nm-thick *a*-plane GaN MQWs with 5.1-nm-thick AlN barriers grown by plasma-assisted molecular-beam epitaxy (PAMBE) on *r*-plane sapphire.[14] Recently, using free-standing *m*-plane GaN substrates, low-temperature ($T$ = 9 K) ISB absorption has been shown at far-infrared wavelengths (47.5-79.5 µm) using *m-G*aN/AlGaN MQWs grown by PAMBE.[15] Room temperature mid-infrared (MIR) ISB absorption in the 4.20 to 4.84 µm range has also been observed recently on *m*-plane GaN/Al$_{0.5}$Ga$_{0.5}$N MQWs grown by metalorganic vapor phase epitaxy (MOVPE).[16] Finally, Pesach *et al.*[17] have demonstrated QW infrared photodetectors (QWIPs) consisting of In$_{0.095}$Ga$_{0.905}$N/Al$_{0.07}$Ga$_{0.93}$N (2.5 nm / 56.2 nm) and In$_{0.1}$Ga$_{0.9}$N/GaN (3 nm /50 nm) MQWs, which displayed photocurrent peaks at 7.5 µm and 9.3 µm, respectively, when characterized at 14 K.

In this paper, we compare GaN/AlN MQWs simultaneously grown on the nonpolar *a*- and *m*-planes as well as on the polar *c*-plane displaying ISB transitions in the short wavelength infrared (SWIR) region. In terms of mosaicity, surface roughness, photoluminescence (PL) linewidth and intensity, and ISB absorption, the best nonpolar results are obtained from *m*-plane structures. With respect to polar structures, the ISB transitions are redshifted, and present similar line widths. Based on these results, we designed a series of *m*-plane GaN/AlGaN MQWs to determine the accessible spectral range in the MIR. These samples show tunable room-temperature ISB absorption from 4.0 to 5.8 µm, where the long-wavelength limit is set by the absorption associated with the second order of the Reststrahlen band in the bulk GaN substrates.

## II. EXPERIMENTAL

The samples were grown by PAMBE at a substrate temperature $T_S$ = 720°C and with a nitrogen-limited growth rate of 0.4 ML/s (≈ 360 nm/h). Growth was performed under the optimum conditions for *c*-plane GaN, i.e. slightly Ga-rich conditions.[8,18,19] For *a*- and *m*-plane GaN/Al(Ga)N heterostructures, the substrates were free-standing semi-insulating GaN sliced along the respective nonpolar surfaces from (0001)-oriented GaN boules synthesized by hydride vapor phase epitaxy (resistivity >10$^6$ Ωcm, dislocation density <5×10$^6$ cm$^{-2}$). For the *c*-plane GaN/Al(Ga)N heterostructures, growth was performed either on 1-µm-thick

AlN-on-sapphire templates (for SWIR structures) or on 4-µm-thick GaN-on-Si(111) templates (for MIR structures), both deposited by MOVPE. The heterostructures were simulated using the Nextnano[3] 8×8 k.p self-consistent Schrödinger-Poisson solver,[20] with the material parameters described by Kandaswamy et al.[8]

The surface morphology of the layers was studied by field-emission scanning electron microscopy (SEM) using a Zeiss Ultra 55 microscope, and by atomic force microscopy (AFM) in the tapping mode using a Dimension 3100 system. The periodicity and structural properties of the MQWs were studied by X-ray diffraction (XRD) using a Seifert XRD 3003 PTS-HR diffractometer with a beam concentrator in front of a Ge(220) 2- or 4-bounce monochromator and a 0.15 degree long plate collimator in front of the detector.

Photoluminescence (PL) spectra were obtained by exciting with a continuous-wave solid-state laser ($\lambda = 244$ nm), with an excitation power around 100 µW focused on a spot with a diameter of ≈100 µm. The emission from the sample was collected by a Jobin Yvon HR460 monochromator equipped with an ultraviolet-enhanced charge-coupled device camera. All PL measurements were performed at 5 K.

Fourier transform infrared spectroscopy (FTIR) was used to probe the ISB absorption using a halogen lamp and a mercury-cadmium-telluride detector incorporated into a Bruker V70v spectrometer. All samples were polished at 45° (bulk GaN or sapphire substrates) or at 30° (Si substrates) to form multipass waveguides allowing 4-5 interactions with the active region. The samples were tested in transmission mode using an MIR polarizer to discern between the transverse-electric (TE) and transverse-magnetic (TM) polarized light. Observation of ISB absorption requires a component of the electric field perpendicular to the QW plane, i.e. TE polarized light is not absorbed.[21] All FTIR measurements were performed at room temperature.

## III. RESULTS

### A. SWIR absorption in GaN/AlN MQWs

To compare the different crystal orientations a series of 40-period GaN/AlN MQWs was grown along the *m*-, *a*- and *c*- crystallographic directions simultaneously. This comparison was possible because each of the three crystallographic planes grows two-dimensionally under Ga-rich conditions in PAMBE.[22–24] These structures were designed to display ISB transitions in the 292-795 meV (1.4-4.2 μm) spectral range. The QWs were doped with Si at a concentration of ≈1×10$^{19}$ cm$^{-3}$. The geometry of the samples and their experimentally-obtained optical properties are summarized in Table I.

To evaluate their structural quality, the surface morphology of the samples was assessed by SEM and AFM, as illustrated in Figure 1 for samples S3a and S3m. On a large scale, the SEM images of the nonpolar samples show smooth surfaces with cracks propagating along the *c* axis (average distance between cracks ≈10 μm), resulting in {11$\bar{2}$0} or {1$\bar{1}$00} facets for *m*- or *a*-oriented samples, respectively. In the polar case, crack propagation occurs when GaN/AlGaN heterostructures are grown under tensile strain. In this case, cracks are isotropically distributed, and present vertical {1$\bar{1}$00} facets.[25,26] Due to the anisotropy of the nonpolar lattices, relaxation along the *c* and *a/m* directions must be analyzed independently. Cracks propagating along the in-plane axis *m* have been described in *a*-AlN grown on *a*-plane 6H-SiC,[27] which was explained as due to the tensile strain along the *c* axis (−1.1% lattice mismatch) in combination with a lack of low-energy slip systems available for plastic relaxation. However, in a highly compressed configuration (with larger mismatch such as *a*-GaN on *r*-sapphire having +1.2% and +16.1% lattice mismatch along *c* and *m*, respectively), cracks are observed to propagate preferentially along the *c* axis.[28,29] In the case of *m*-AlGaN, cracks propagating along the in-plane axis *a* have been reported.[30,31] For the samples in this study, it appears that these defects do not develop during the growth, but instead during the cooling process as a result of the temperature-dependent GaN/AlN lattice mismatch.[32] On the AFM scale, the root-mean-square (rms) surface roughness measured in images of an area of 5×5 μm$^2$ was 1.1±0.2 nm, 2.0±0.6 nm, and 3.7±1.2 nm for *c*-, *m*-, and *a*-plane samples, respectively, i.e. *m*-plane growth systematically resulted in smoother surfaces than *a*-plane growth.

The periodicity and strain state of the samples were analyzed by XRD. Figure 2 presents the θ–2θ scans of the ($3\bar{3}00$) reflection of samples S1m, S2m, and S3m, and the ($11\bar{2}0$) reflection of samples S1a and S2a. Table I summarizes the MQW period extracted from the inter-satellite distance in the XRD measurements. The full width at half maximum (FWHM) of the rocking curves were measured for the substrate and the MQW zero-order reflection with $\phi = 0°$ and $\phi = 90°$ ($\Delta\omega_c$ and $\Delta\omega_a$, respectively for the *m*-oriented samples, and $\Delta\omega_c$ and $\Delta\omega_m$, respectively for the *a*-oriented sample), which provides information on the sample mosaicity in the *c* and *a* directions, respectively for the *m*-oriented samples, and in the *c* and *m* directions, respectively for the *a*-oriented samples. Comparing the values in Table I, it appears that the *m*-plane MQWs exhibit better crystalline quality than the *a*-plane structures.

To assess the MQW strain state, reciprocal space maps were measured. Figure 3 illustrates the results for sample S2m, where the ($3\bar{3}00$), ($3\bar{3}02$), and ($3\bar{2}\bar{1}0$) reflections were considered. The reciprocal space is presented using the GaN substrate as a reference. The shift in q(0001) (projection of the reciprocal space vector along [0001]) of the MQW ($3\bar{3}00$) reflection with respect to the substrate (see Fig. 3(b)) reveals a tilt of the epitaxial structure. The tilt angles towards the in-plane directions ($\delta\omega_c$ and $\delta\omega_a$ for *m* oriented samples, and $\delta\omega_c$ and $\delta\omega_m$ for *a* oriented samples) are summarized in Table I. Taking the measured tilt into account, the strain states along the *a*, *m* and *c* axis can be described as $\varepsilon_p = \frac{p - p_r}{p_r}$, where $\varepsilon_p$ is the strain along the axis *p* (*a*, *m* or *c*), *p* is the measured lattice parameter along this axis, and $p_r$ is the theoretical value of *p* assuming that the structure is relaxed. Using the lattice parameters of Vurgaftman *et al.*[34] and Wright *et al.*[35] ($a_{GaN}$ = 3.1891 Å; $a_{AlN}$ = 3.112 Å; $c_{GaN}$ = 5.1850 Å; $c_{AlN}$ = 4.980 Å), the lattice mismatch between AlN and GaN is –2.4% in the *a* and *m* directions and –3.9% in the *c* direction. The larger mismatch along *c* explains the larger tilt towards this direction (0.05° to 0.29°). This tilt is a way to relax the in-plane lattice mismatch, and thus to reduce the number of dislocations necessary to release the strain.[33]

Figure 4 presents the values of strain extracted from the reciprocal space maps, compared with the in-plane lattice mismatch between the relaxed MQWs (considered as a relaxed AlGaN alloy with the average Al composition of the structure) and the GaN substrates. Due to the lattice mismatch, all the structures undergo in-plane tensile strain, and as a result of Hooke's law, they are compressively strained in the growth direction. In the case of *m*-oriented samples, all the MQWs are about 50% relaxed along the in-plane *a* axis, whereas almost full relaxation is observed along *c*.

The PL spectra of all the samples were measured at low temperature ($T$ = 5 K), with the results in terms of emission wavelength and intensity summarized in Table I. As an illustration, Figure 5(a) shows the spectra of samples S2c, S2m and S2a. For all samples, the *c*- and *a*- orientations systematically lead to broader emission peaks than those measured for the *m*-orientation. In addition, the PL from *m*-plane samples is twice as intense as that from *a*-plane, and more than twenty times as intense as that from *c*-plane QWs.

In Figure 5(b), the PL peak emission energies are compared with theoretical calculations assuming that the in-plane lattice parameters correspond to those of an AlGaN ternary alloy with an Al composition equal to the average Al content of the MQW. For nonpolar samples, their luminescence is systematically above the GaN band gap, supporting the absence of internal electric field. For the *c*-plane samples, the emission energy shifts below the GaN band gap when increasing the well width. In general, the emission energies are in agreement with the theoretical calculations. The deviation from the calculations observed for S1 is attributed to carrier localization in thickness fluctuations in such small QWs (the thickness of a GaN monolayer being ≈ 0.25 nm).

The ISB absorption in the SWIR range was measured at room temperature by FTIR spectroscopy. To identify the ISB transition in the samples, the TE transmission spectra were divided by the respective TM transmission spectra, and the results are presented in Figure 6(a) and Table I. As expected, the absorption is red-shifted when decreasing the QW width. In the case of nonpolar MQWs, the absence of the internal electric field results in a red shift of the ISB energy, in comparison to *c*-plane structures, where the triangular potential profile in the wells contributes to the separation of the quantized electron levels. A similar result was observed in the case of semipolar $(11\bar{2}2)$ MQWs due to the reduction of spontaneous and piezoelectric polarizations.[36] Nonpolar *m*-plane samples exhibit an absorption linewidth similar to that of polar MQWs absorbing at the same wavelengths. In contrast, the TM polarized absorption of *a*-plane sample S2a undergoes a significant broadening and deviation from the calculations, and no ISB absorption was observed for sample S3a.

In summary, ISB transitions in *m*-oriented GaN/AlN MQWs can cover the SWIR spectral range with performance comparable to polar MQWs, with the advantage of design simplicity in a geometry with square potential band profiles. Furthermore, *m*-plane structures display better results in terms of mosaicity, surface roughness, PL linewidth and intensity, and ISB absorption than those obtained when growing on the nonpolar *a* plane.

## B. MIR absorption in GaN/AlGaN MQWs

In a second stage, we have analysed the possibility of covering the MIR spectral region with nonpolar QWs. Based on the previous results, only the *m* crystallographic orientation was considered. The QWs were enlarged to achieve the desired spectral shift, and the AlN barriers were replaced by the ternary alloy AlGaN with a twofold purpose: reducing the lattice mismatch in the MQW and approaching the excited level in the QW to the continuum, to mimic the band diagram of a QWIP. The barriers were chosen to be 22.6 nm thick, in order to prevent coupling between QWs even in the largest QWs. Four *m*-plane structures were designed to display ISB transitions between the ground conduction band level and the first excited level ($e_1 \rightarrow e_2$) in the 186-356 meV (3.4-6.7 µm) range, using the QW thicknesses and Al contents in the barriers summarized in Table II. Note that the use of bulk GaN as a substrate sets an additional limit for characterization. Even though the GaN Reststrahlen band spans from 9.6 µm to 19 µm, absorption in the range of 6.7 µm to 9 µm has been observed in bulk GaN substrates with carrier concentrations $<10^{16}$ cm$^{-3}$, and was attributed to the second harmonic of the Reststrahlen band.[37–39] Figure 7 shows the band diagrams of the *m*-plane MQWs, together with those of structures with the same dimensions but grown along the *c* direction. In the case of the *c*-oriented MQWs studied in this work, characterization of ISB absorption in the spectral range between 6.7 µm and 9 µm is possible due to the use of floating-zone silicon substrates, as previously demonstrated.[40]

A series of 50-period GaN/Al$_x$Ga$_{1-x}$N MQWs was grown along the *m*- and *c*-crystallographic directions simultaneously, following the designs in Table II. As a first evaluation of the structural quality, the surface morphology was assessed by AFM and SEM, as illustrated in Figure 8 for sample S4m. Similar to the SWIR samples, SEM images of the nonpolar samples reveal cracks propagating along the *c*-axis. However, the distance between cracks increased to ≈15-30 µm. At the AFM scale, all the nonpolar samples in Table II present similar morphology: large-scale (5×5 μm² to 10×10 μm² images) roughness in the range of 7-15 nm, whereas at a smaller scale (1×1 μm² images) the surfaces are smooth, with rms roughness in the 1-2 nm range.

The periodicity, strain state and mosaicity of sample S7m were analyzed by XRD. To assess the MQW strain state, we measured various reciprocal space maps for sample S7m (the error bars of this technique were too large to extract reliable conclusions in samples containing lower Al content). The extracted strain states are $\varepsilon_m = 0.03\pm0.15\%$, $\varepsilon_a = -0.43\pm0.40\%$, and $\varepsilon_c = -0.27\pm0.40\%$. Compared to the relaxed lattice mismatch between the GaN/Al$_{0.44}$Ga$_{0.56}$N MQW and the GaN substrate (–0.98% in the *a* and *m* directions and –1.6% in the *c* direction), they point to a certain relaxation in spite of the large error bars of the measurement. The FWHM of the rocking curves of the MQW reflection were $\Delta\omega_c = 0.28°$ and $\Delta\omega_a = 0.22°$, pointing to a significant improvement of the MQW crystalline quality with respect to the GaN/AlN QWs (see Table I) owing to the reduced lattice mismatch.

The PL spectra of all the samples were measured at low temperature, as illustrated by Figure 9(a). In Figure 9(b), the PL emission energies are compared with theoretical calculations as a function of the QW width. For the *c*-plane samples, the luminescence is systematically below the GaN bandgap due to the internal electric field, and it exhibits superimposed oscillations due to Fabry-Perot interferences. For nonpolar samples, the emission remains above the GaN band gap energy. In both cases, decreasing the QW width leads to a red shift of the PL energy, with emission energies in agreement with the calculations.

The ISB absorption in the MIR range was measured at room temperature by FTIR spectroscopy. To identify the ISB transition in the samples, the substrate transmission spectrum was divided by the respective TM transmission spectra, with the results displayed in Figures 10(a) and (b). In nonpolar structures, increasing the QW width leads to a red shift of the ISB energies from 308 to 213 meV (4.0 to 5.8 µm), in agreement with calculations as shown in Figure 10(c). The deviation observed in the sample with the largest QWs (calculated transition at 186 meV) is attributed to the proximity of the second order of the Reststrahlen band at 184 meV (6.7 µm), which sets the onset of substrate absorption [shadowed area in Figures 10(a) and (c)]. For all polar samples, two absorption peaks are observed. The peak at lower energy corresponds to the ($e_1 \rightarrow e_2$) transition, whereas the higher energy peak is assigned to ISB transitions involving upper states ($e_1 \rightarrow e_3$, $e_1 \rightarrow e_4$), as previously observed in GaN/AlN QWs.[41] In symmetric structures, the $e_1 \rightarrow e_3$ transition is forbidden due to parity, whereas $e_1 \rightarrow e_4$ is allowed. However, both transitions are possible in asymmetric polar QWs, and the second peak might hence correspond to the combination of both transitions. In this series of samples, the absorption in *m*- and *c*-oriented MQWs is located in the same spectral range, both theoretically and experimentally. This coincidence is due to the choice of the Al content in the barriers, which determines the energetic location of $e_2$ in the *c*-plane structures. Increasing the Al content of the barriers would introduce only slight corrections to ($e_1 \rightarrow e_2$) in *m*-plane MQWs, but it would induce a major blue shift of this transition in the *c*-plane MQWs due to the internal electric field.

**IV CONCLUSIONS**

In summary, we have shown room temperature SWIR ISB absorption in a series of nonpolar *a*- and *m*- plane and polar *c*-plane GaN/AlN MQWs with various QW thicknesses. Comparing the two nonpolar crystallographic planes, the best results in terms of mosaicity, surface roughness, PL linewidth and intensity, and ISB absorption were obtained for *m*- oriented samples. We have demonstrated that ISB transitions in *m*-GaN/AlN MQWs can cover the whole SWIR spectrum (1.5-2.9 µm) with performance comparable to polar MQWs and with the advantage of design simplicity. The ISB absorption is systematically red shifted with respect to polar structures with the same geometry due to the triangular potential profile induced by the internal electric field. Drawing from the experience in the SWIR range, we have designed a series of *m*-plane GaN/AlGaN MQWs with ternary barriers and with larger QWs, to shift the ($e_1 \rightarrow e_2$) ISB energy towards the MIR. We have demonstrated experimentally that the ISB absorption in these *m*-plane samples can be tuned in the range of 4.0-5.8 µm, the longer wavelength limit being set by the second order of the GaN Reststrahlen band when using bulk substrates.

**ACKNOWLEDGEMENTS.** This work is supported by the EU ERC-StG "TeraGaN" (#278428) project.

## TABLES

**Table I:** Structural and optical characteristics of the GaN/AlN MQW samples on GaN substrates: QW thickness ($t_{QW}$) (barrier thickness is 3.6 nm for all samples); MQW period measured by XRD; broadening of the ω-scan of the ($3\bar{3}00$) XRD reflection in the *c* and *a* directions ($\Delta\omega_c$ and $\Delta\omega_a$, respectively) for *m*-oriented samples and substrates, and broadening of the ω-scan of the ($11\bar{2}0$) XRD reflection in the *c* and *m* directions ($\Delta\omega_c$ and $\Delta\omega_m$, respectively) for *a*-oriented samples and substrates; tilt between the MQW and the GaN substrate towards the *c* and *a* directions ($\delta\omega_c$ and $\delta\omega_a$, respectively) for *m*-oriented samples, and towards the *c* and *m* directions ($\delta\omega_c$ and $\delta\omega_m$, respectively) for *a*-oriented samples; strain state in the 3 perpendicular directions *m*, *a* and *c* ($\varepsilon_m$, $\varepsilon_a$, and $\varepsilon_c$ respectively); PL peak energy and intensity normalized with respect to S1m; simulated and measured ISB transition energy.

| Sample | $t_{QW}$ (nm) | XRD Period (nm) | XRD FWHM MQW (°) | XRD FWHM GaN (°) | Tilt MQW/GaN (°) | Strain (%) | PL peak energy (eV) [normalized intensity] | Simulated / Measured ISB transition (meV) |
|---|---|---|---|---|---|---|---|---|
| S1m | 1.5 | 5.1 | $\Delta\omega_c = 0.33$<br>$\Delta\omega_a = 0.39$ | $\Delta\omega_c = 0.028$<br>$\Delta\omega_a = 0.028$ | $\delta\omega_c = 0.20$<br>$\delta\omega_a = 0.04$ | $\varepsilon_m = 0.38\pm0.15$<br>$\varepsilon_a = -1.03\pm0.40$<br>$\varepsilon_c = -0.36\pm0.40$ | 3.8 [1] | 712 / 799 |
| S2m | 2.3 | 5.9 | $\Delta\omega_c = 0.44$<br>$\Delta\omega_a = 0.30$ | $\Delta\omega_c = 0.037$<br>$\Delta\omega_a = 0.040$ | $\delta\omega_c = 0.08$<br>$\delta\omega_a = 0.015$ | $\varepsilon_m = 0.44\pm0.15$<br>$\varepsilon_a = -0.50\pm0.40$<br>$\varepsilon_c = -0.17\pm0.40$ | 3.8 [0.94] | 437 / 578 |
| S3m | 3.1 | 6.7 | $\Delta\omega_c = 0.45$<br>$\Delta\omega_a = 0.29$ | $\Delta\omega_c = 0.026$<br>$\Delta\omega_a = 0.032$ | $\delta\omega_c = 0.05$<br>$\delta\omega_a = 0.01$ | $\varepsilon_m = 0.41\pm0.15$<br>$\varepsilon_a = -0.63\pm0.40$<br>$\varepsilon_c = -0.23\pm0.40$ | 3.7 [0.58] | 296 / 425 |
| S1a | 1.5 | 5.1 | $\Delta\omega_c = 0.28$<br>$\Delta\omega_m = 0.72$ | $\Delta\omega_c = 0.030$<br>$\Delta\omega_m = 0.030$ | $\delta\omega_c = 0.16$<br>$\delta\omega_m = 0.00$ | $\varepsilon_a = 0.33\pm0.15$<br>$\varepsilon_m = -0.56\pm0.40$<br>$\varepsilon_c = -0.62\pm0.40$ | 3.9 [0.56] | 712 / 815 |
| S2a | 2.3 | 5.9 | $\Delta\omega_c = 0.53$<br>$\Delta\omega_m = 0.40$ | $\Delta\omega_c = 0.019$<br>$\Delta\omega_m = 0.023$ | $\delta\omega_c = 0.29$<br>$\delta\omega_m = 0.015$ | $\varepsilon_a = 0.32\pm0.15$<br>$\varepsilon_m = -0.42\pm0.40$<br>$\varepsilon_c = -0.09\pm0.40$ | 3.7 [0.24] | 431 / 755 |
| S3a | 3.1 | 6.7 | -- | -- | -- | -- | 3.7 [0.11] | 296 / -- |
| S1c | 1.5 | 5.1 | -- | -- | -- | -- | 3.7 [0.014] | 814 / 815 |
| S2c | 2.3 | 5.9 | -- | -- | -- | -- | 3.2 [0.008] | 657 / 731 |
| S3c | 3.1 | 6.7 | -- | -- | -- | -- | 3.0 [0.003] | 603 / 624 |

**Table II:** Structural and optical characteristics of the GaN/AlGaN MQW samples: QW thickness ($t_{QW}$) (barrier thickness is 22.6 nm for all samples); MQW period measured by XRD; Al composition in the barrier ($x_B$); PL peak energy; simulated and measured first ($e_1 \rightarrow e_2$), second ($e_1 \rightarrow e_3$) and third ($e_1 \rightarrow e_4$) ISB transition energies. Samples S4, S5, and S7 were Si-doped with $[Si] \approx 2 \times 10^{19}$ cm$^{-3}$. Sample S6 was doped with $[Si] \approx 8 \times 10^{18}$ cm$^{-3}$. (*) Thickness extrapolated from XRD measurements of other samples in the same series.

| Sample | $t_{QW}$ (nm) | XRD Period (nm) | $x_B$ (%) | PL peak energy (eV) | Simul. ($e_1 \rightarrow e_2$) / Meas. ISB transition (meV) | Simul. ($e_1 \rightarrow e_3$, $e_1 \rightarrow e_4$) / Meas. ISB transition (meV) |
|---|---|---|---|---|---|---|
| S4m | 3.1 | 25.7 (*) | 26 | 3.60 | 186 / 222 | -- |
| S5m | 2.8 | 25.4 (*) | 31 | 3.64 | 223 / 213 | -- |
| S6m | 2.5 | 25.1 (*) | 35 | 3.39 | 261 / 251 | -- |
| S7m | 2.0 | 24.6 (*) | 44 | 3.40 | 356 / 308 | -- |
| S4c | 3.1 | 25.7 (*) | 26 | 3.61 | 162 / 188 | 227, 256 / 270 |
| S5c | 2.8 | 25.4 | 31 | 3.68 | 200 / 209 | 264, 294 / 319 |
| S6c | 2.5 | 25.1 | 35 | 3.41 | 226 / 241 | 292, 323 / 326 |
| S7c | 2.0 | 24.6 | 44 | 3.46 | 290 / 286 | 358, 387 / 378 |

**Figure captions**

**Figure 1.** SEM and AFM images of samples (a) S3a and (b) S3m.

**Figure 2.** XRD θ-2θ scans of the ($3\bar{3}00$) reflection of samples S1m, S2m, and S3m, and the ($11\bar{2}0$) reflection of samples S1a and S2a. The corresponding QW thicknesses are indicated at the right side of the figure.

**Figure 3.** Reciprocal space maps of sample S2m around asymmetric reflections (a) ($3\bar{3}02$) with the *c*-axis in the diffraction plane and (c) ($3\bar{2}\bar{1}0$) with the *a*-axis in the diffraction plane, and symmetric reflection (b) ($3\bar{3}00$) oriented along *c*. (cps= counts per second)

**Figure 4.** Strain state of the MQWs extracted from XRD measurements. (a) In-plane lattice parameter *a* (*m* samples) or *m* (*a* samples). (b) In-plane lattice parameter *c*. (c) Out-of-plane lattice parameter. Positive (negative) values of strain correspond to compressive (tensile) strain. Dashed lines indicate the lattice mismatch between a relaxed AlGaN layer with the average Al concentration of the MQW and the GaN substrate.

**Figure 5.** (a) PL spectra of samples measured at low temperature ($T = 5$ K). (b) PL peak energies as a function of the QW width. Error bars correspond to the FWHM of the PL peaks. Solid lines are theoretical calculations assuming that the in-plane lattice parameters of the MQWs correspond to those of a relaxed AlGaN alloy with the average Al composition of the structure. Dashed lines mark the location of the GaN band gap.

**Figure 6.** (a) TM-polarized ISB absorption of the samples in Table I measured at room temperature. Data are normalized and vertically shifted for clarity. The corresponding QW thicknesses are indicated on the right side. (b) ISB energies as a function of the QW width. Solid lines correspond to theoretical simulations assuming that the in-plane lattice

parameters of the MQWs correspond to those of a relaxed AlGaN alloy with the average Al composition of the structure.

**Figure 7.** Conduction band diagram with first four energy levels and electron wavefunctions of a QW in the center of the active region of samples (a) S4m, (b) S7m, (c) S4c, and (d) S7c.

**Figure 8.** Typical (a) SEM and (b) AFM images of the GaN/AlGaN MQWs in Table II. Measurements correspond to sample S4m.

**Figure 9.** (a) PL spectra of the *m*- and *c*-plane GaN/AlGaN MQWs in Table II measured at low temperature. Data are normalized and vertically shifted for clarity. The corresponding QW thicknesses are indicated on the left side. (b) PL peak energies as a function of the QW width. Error bars correspond to the FWHM of the PL emission. Solid lines are theoretical calculations of the band-to-band transition assuming that the in-plane lattice parameters of the MQWs correspond to those of a relaxed AlGaN alloy with the average Al composition of the structure. Dashed lines mark the location of the GaN band gap.

**Figure 10.** TM-polarized ISB absorption spectra for (a) the m-plane and (b) c-plane GaN/AlGaN MQWs in Table II measured at room temperature. Data are normalized and vertically shifted for clarity. The corresponding QW thicknesses are indicated on the right side. ISB energies as a function of QW width for all (c) m-plane and (d) c-plane samples. Solid lines are theoretical calculations assuming that the in-plane lattice parameters of the MQWs correspond to those of an AlGaN alloy with the average Al composition of the structure. Shadowed areas in graphs (a) and (c) mark the second order of the Reststrahlen band of GaN.

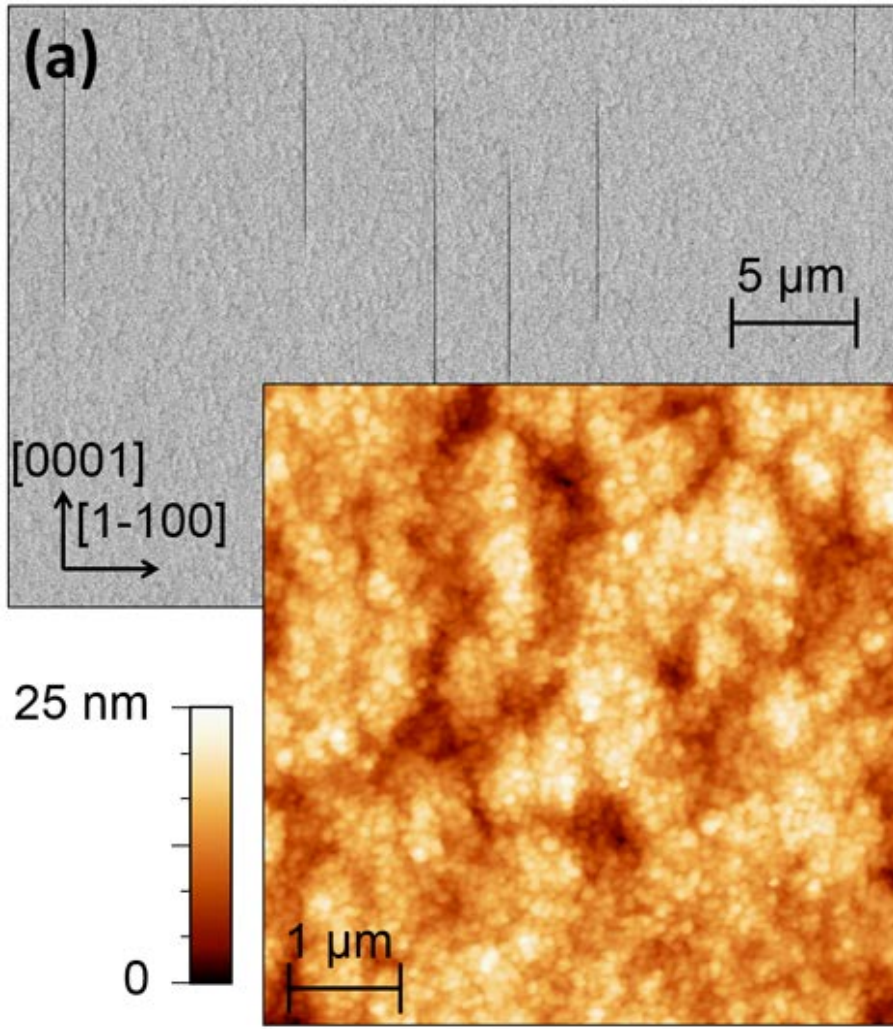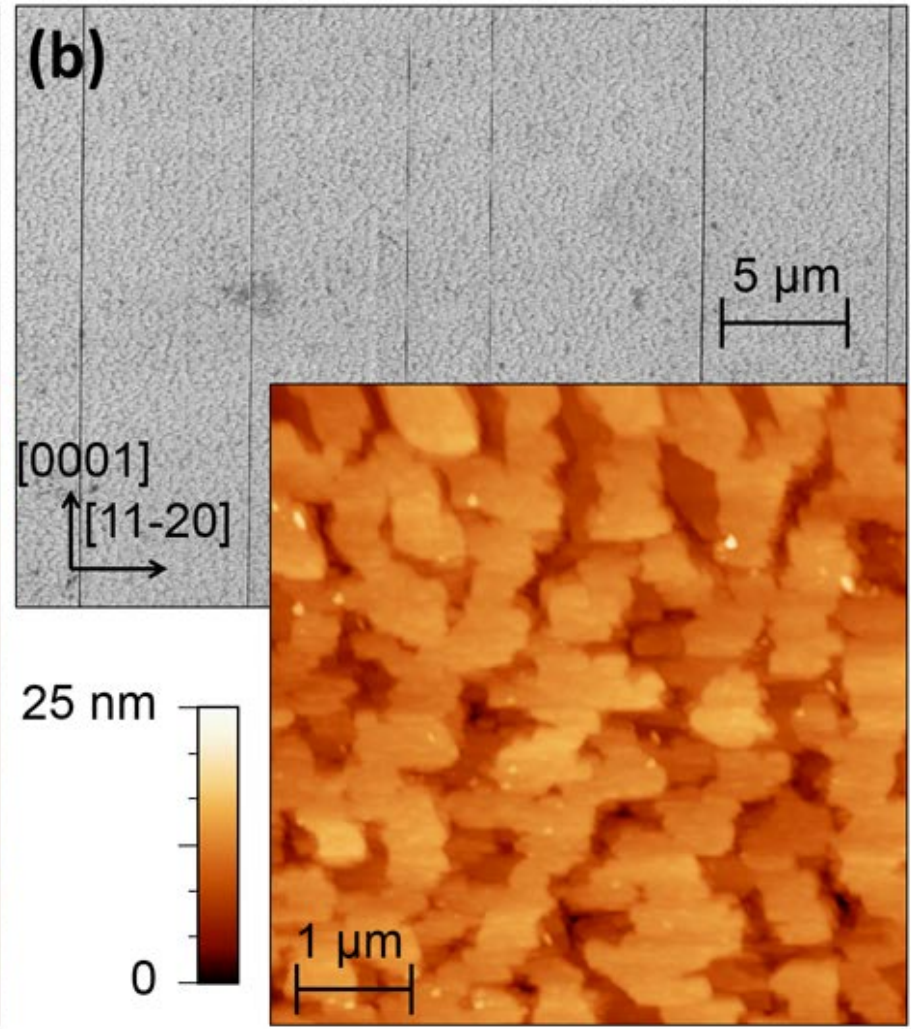

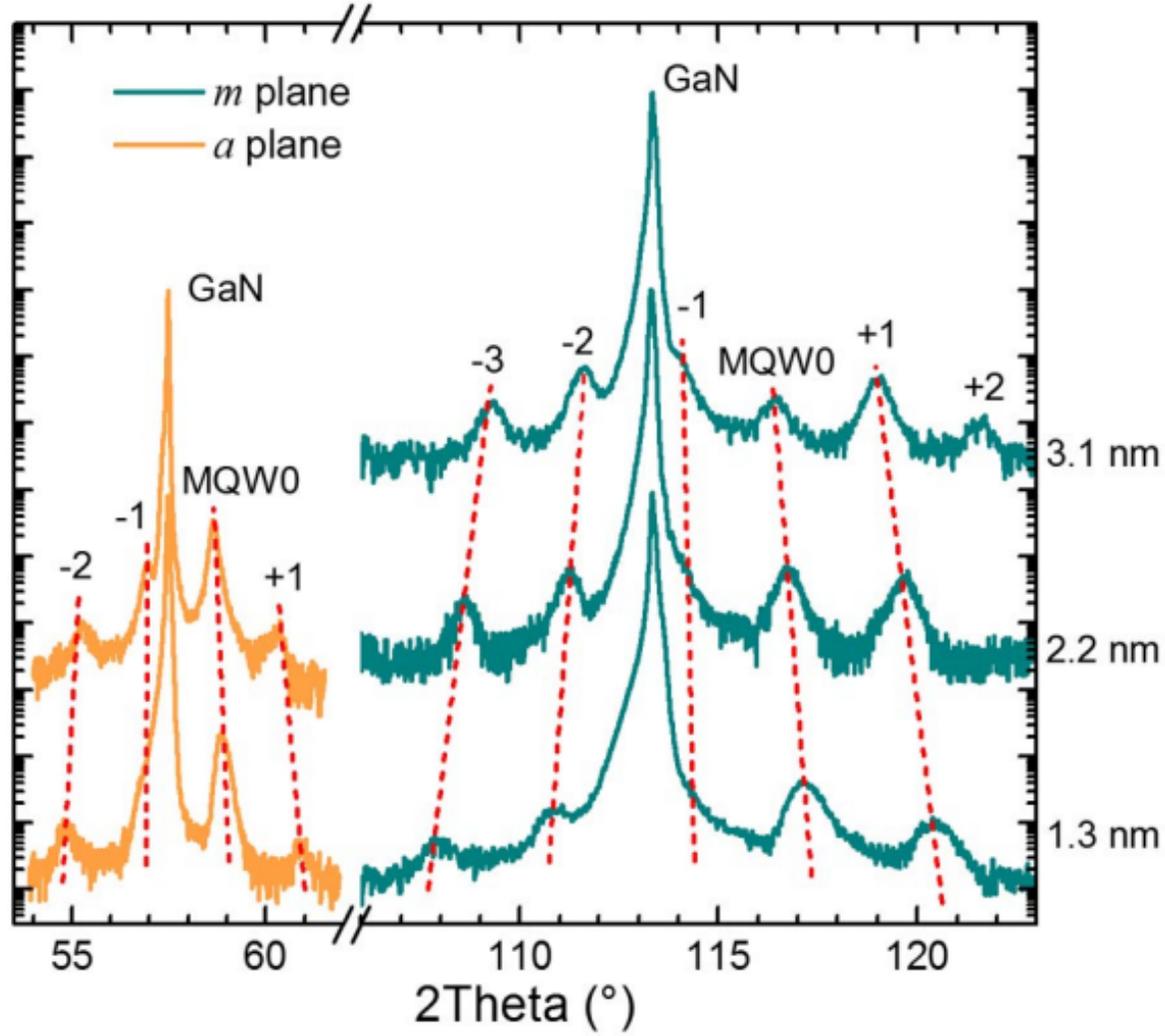

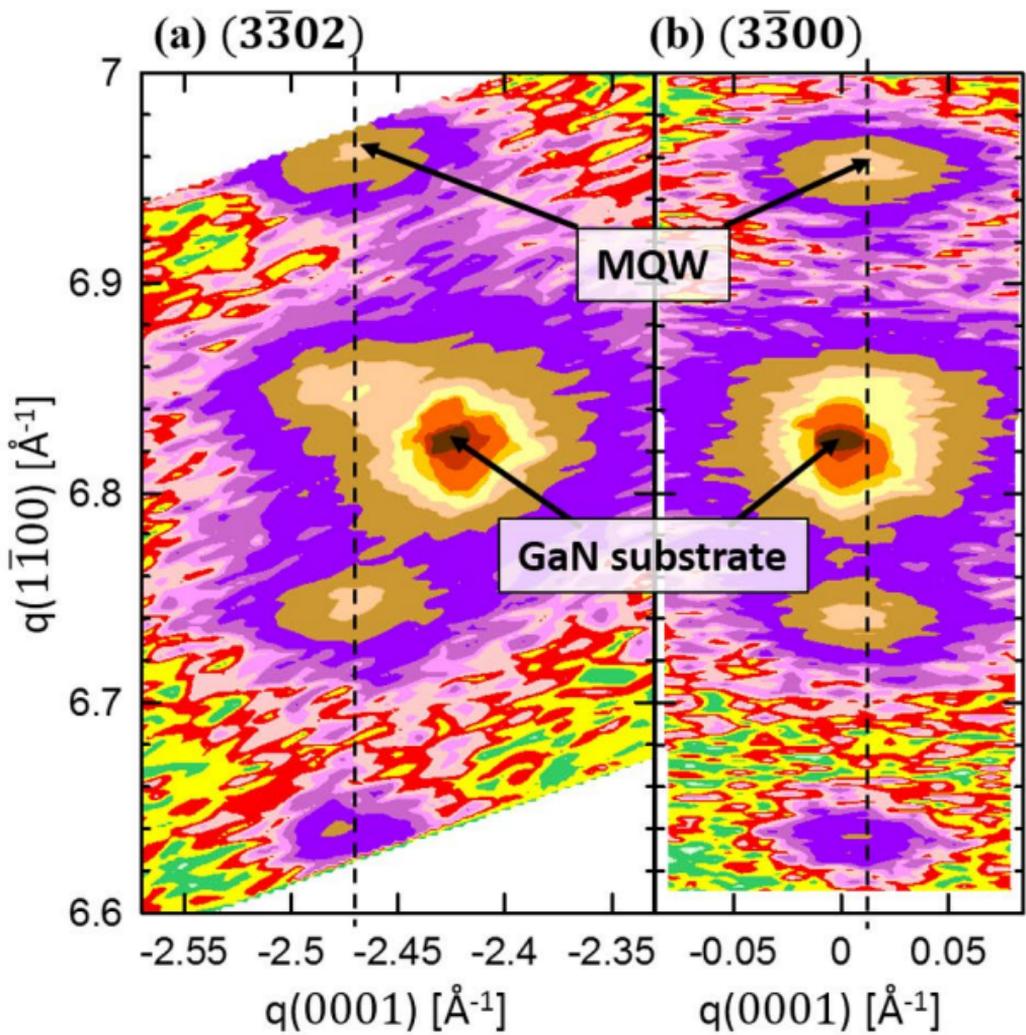
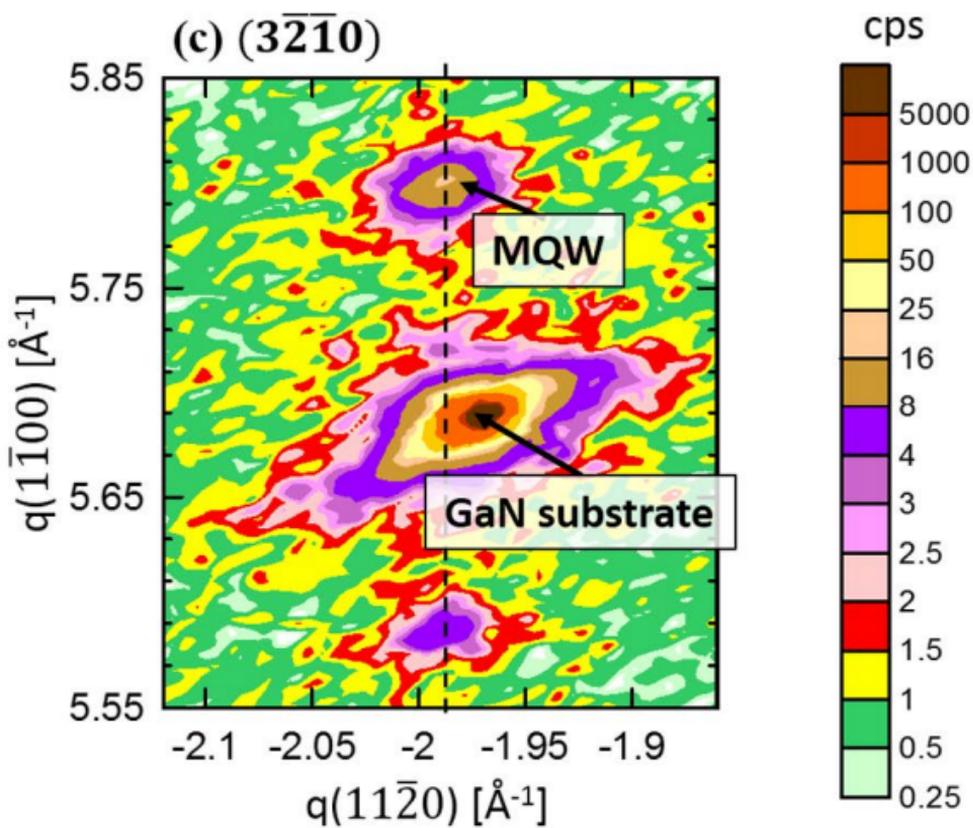

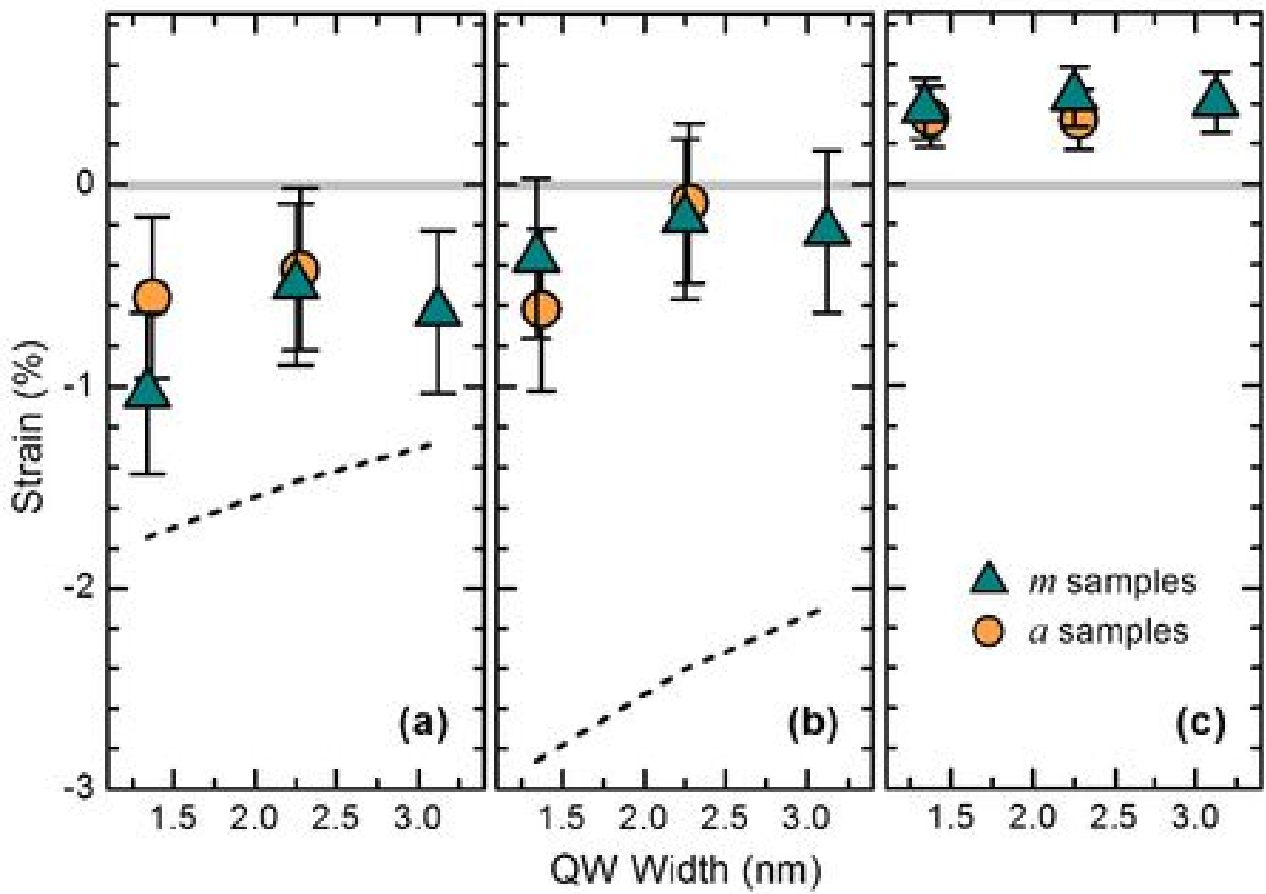

(a) (b) (c)

*m* samples
*a* samples

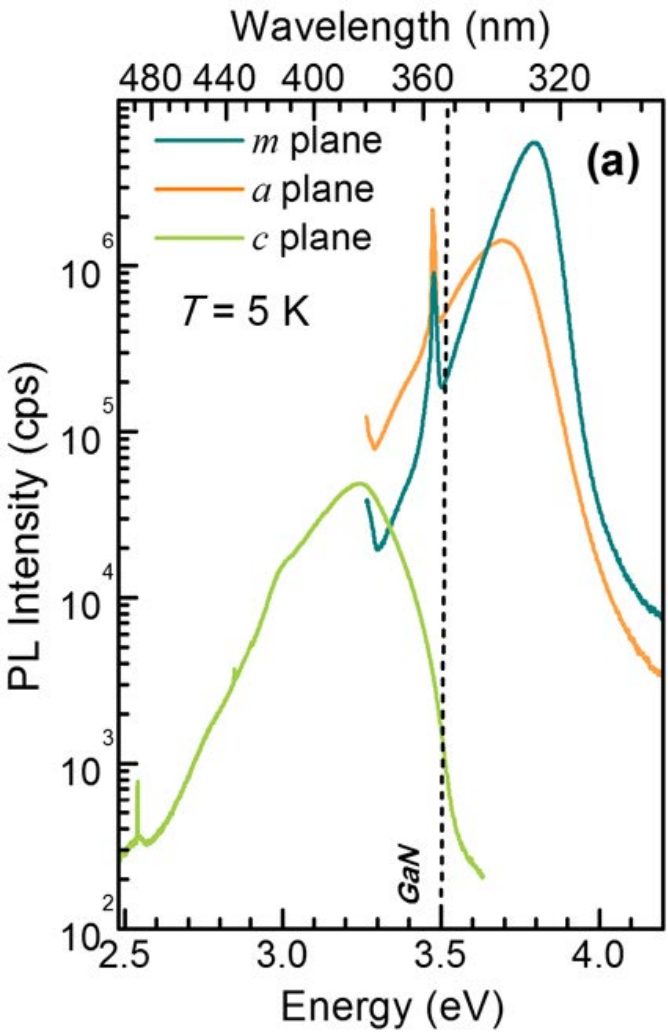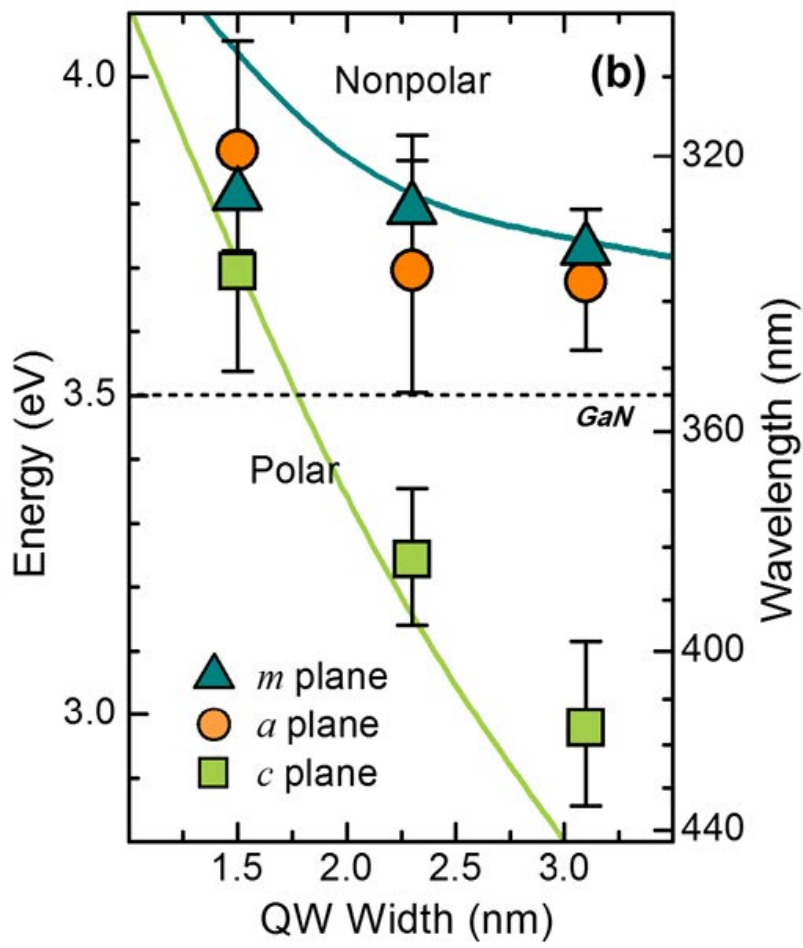

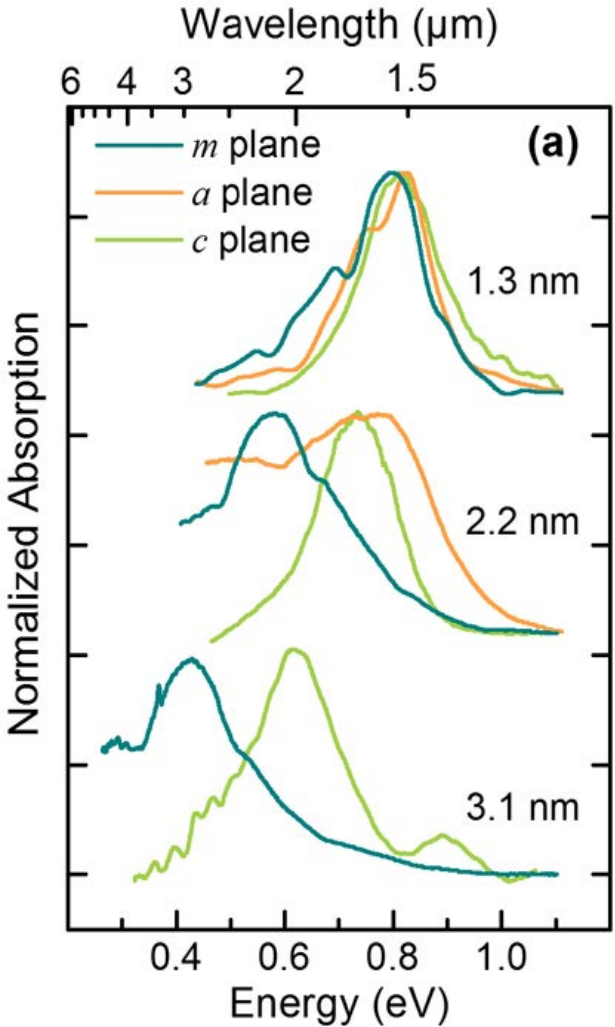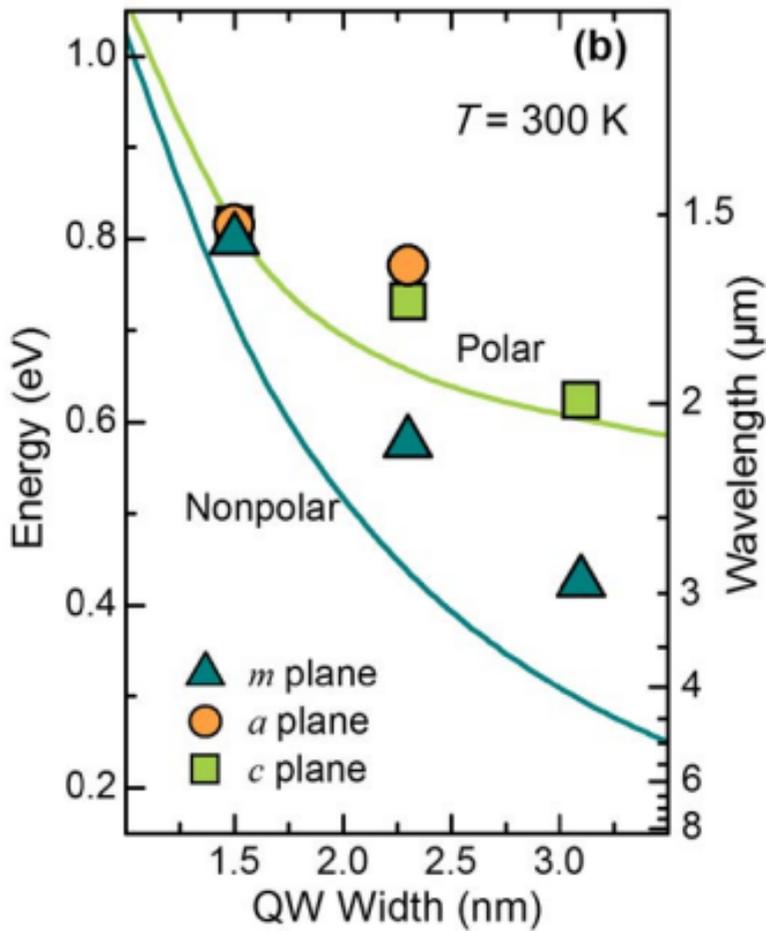

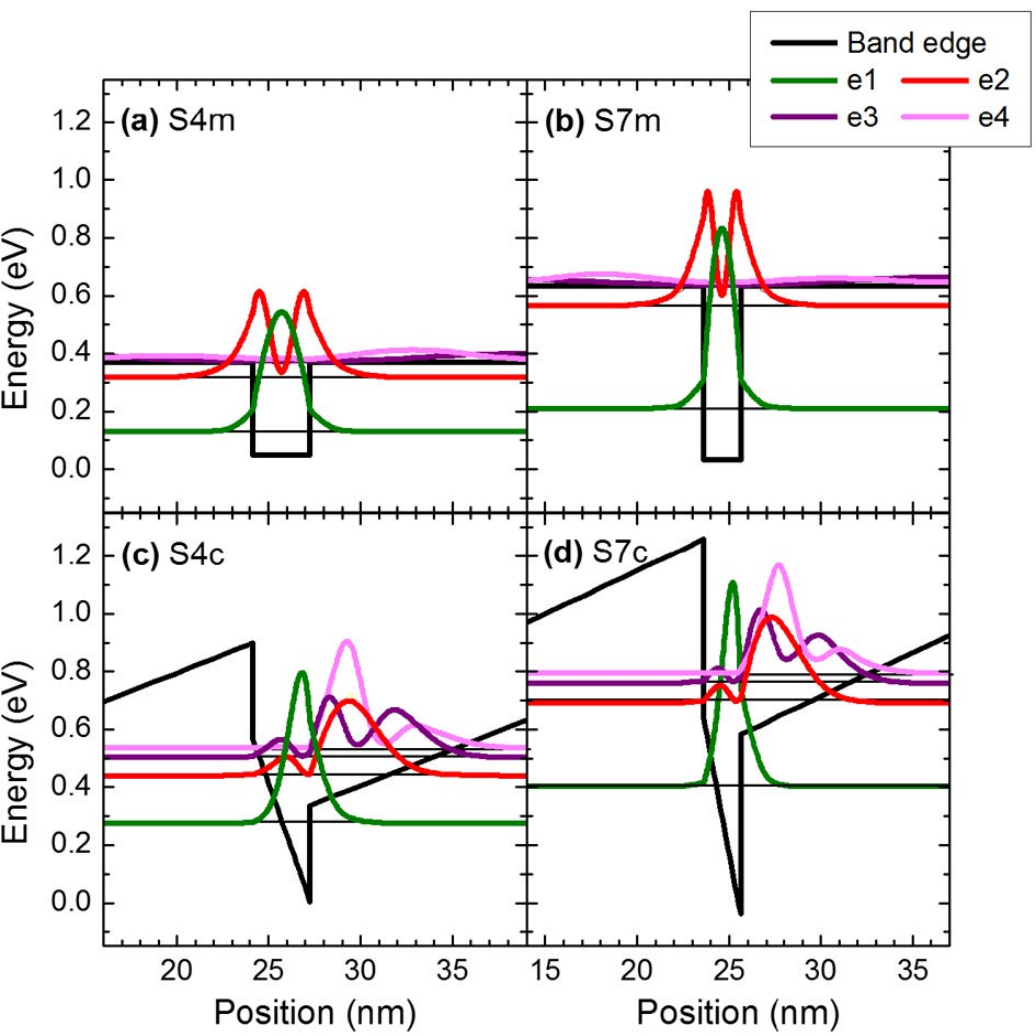

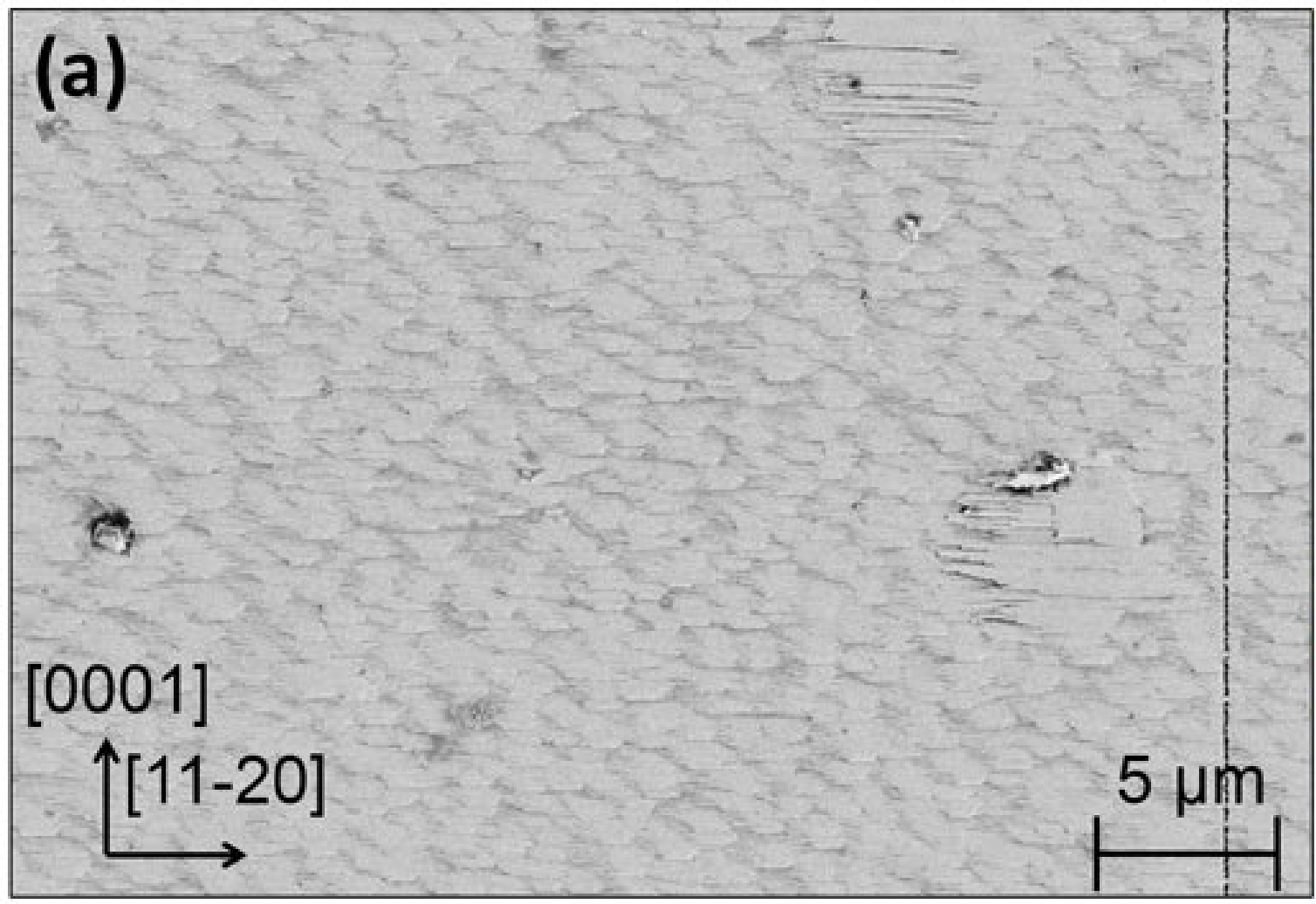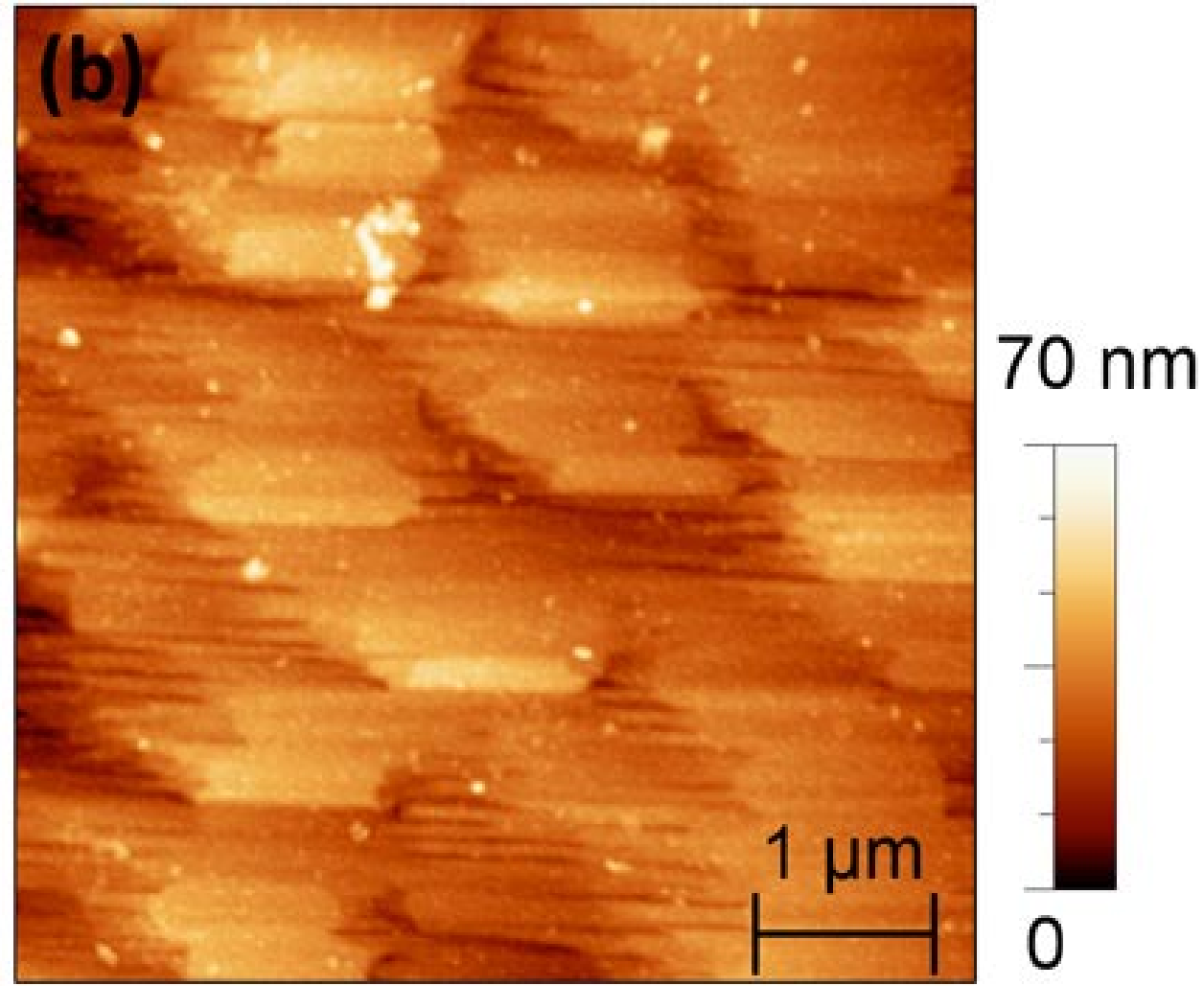

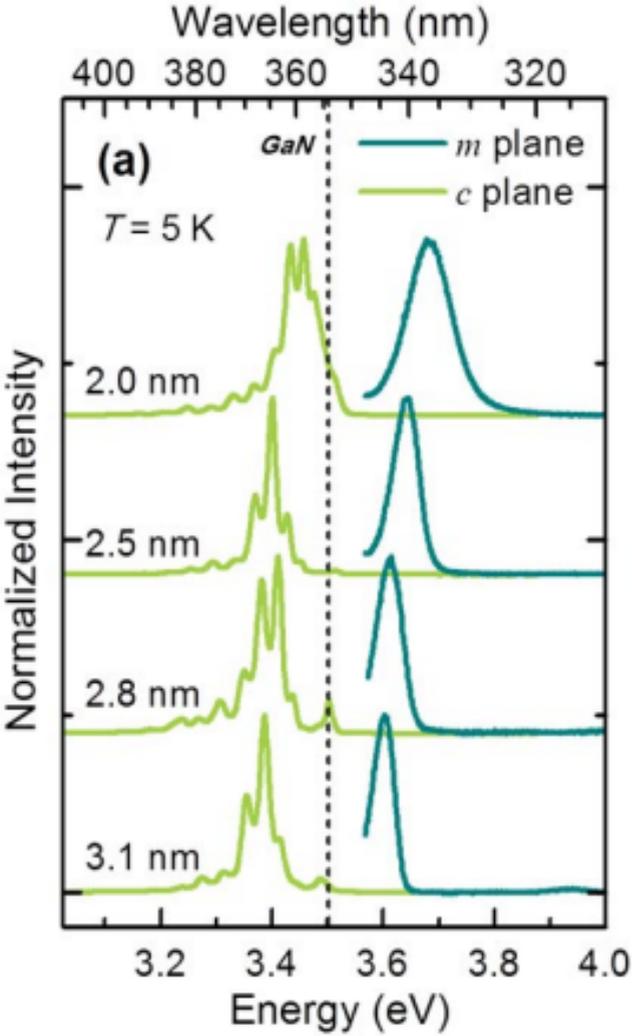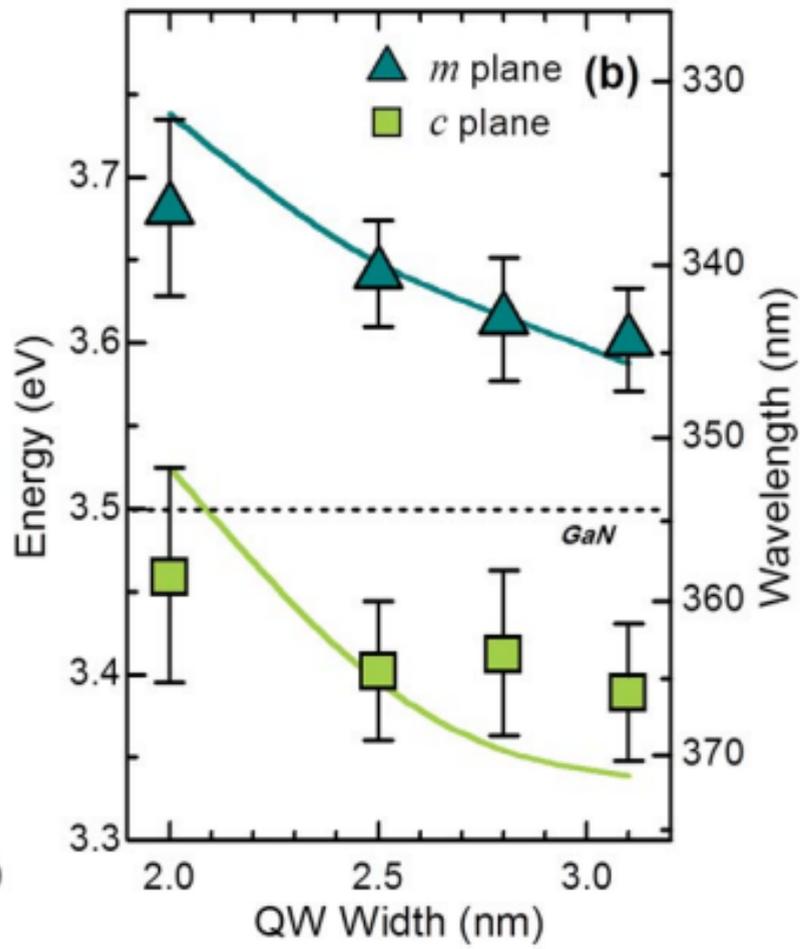

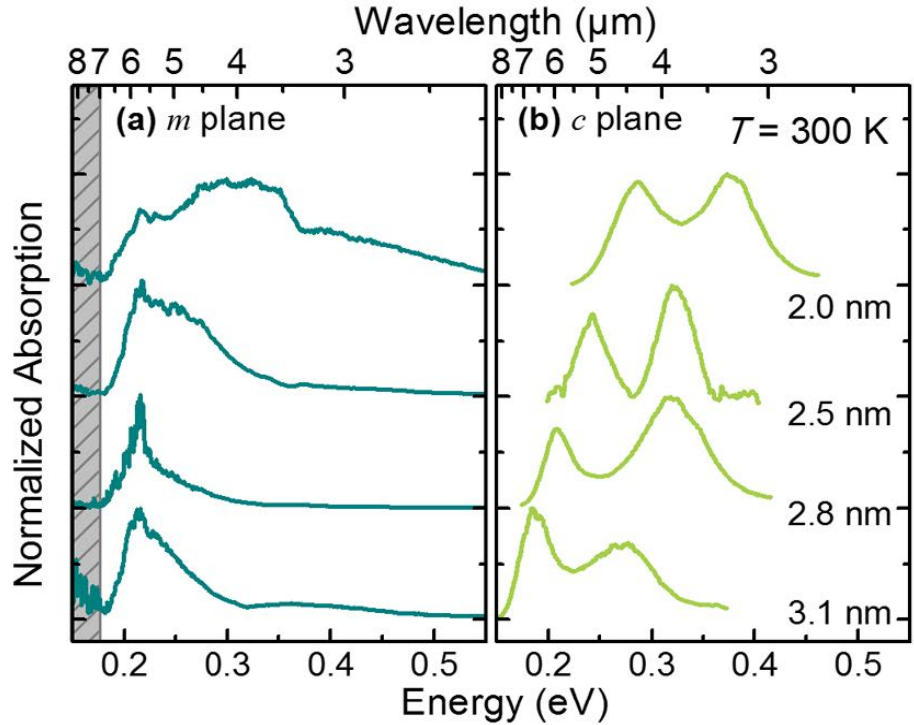